\documentclass[referee]{aa} 

\usepackage{graphicx}
\usepackage{txfonts}
\begin{document} 


   \title{Evolutionary status of isolated B[e] stars}


   \author{Chien-De Lee
          \inst{1}
          \and
          Wen-Ping Chen
          \inst{1,2}
          \and
          Sheng-Yuan Liu
          \inst{3}
          }

   \institute{Graduate Institute of Astronomy, National Central University, 300 Jhongda 
Road, Jhongli 32001, Taiwan
             \email{cdlee@astro.ncu.edu.tw}
         \and
             Department of Physics, National Central University, 300 Jhongda Road, Jhongli 32001, Taiwan
         \and Institute of Astronomy and Astrophysics, Academia Sinica, Taipei 10617, Taiwan
             }


  \abstract
  {} 
{
We study a sample of eight B[e] stars with uncertain evolutionary status
to shed light on the origin of their circumstellar dust.
}
{
We performed a diagnostic analysis on the spectral energy distribution beyond
infrared wavelengths, and conducted a census of neighboring region of each target to ascertain its
evolutionary status.
}
{
In comparison to pre-main sequence Herbig stars, these B[e] stars show equally substantial excess
emission in the near-infrared, indicative of existence of warm dust, but much reduced excess 
at longer wavelengths, so the dusty envelopes should be compact in size.  Isolation from 
star-forming regions excludes the possibility of their pre-main sequence status.  Six of 
our targets, including HD\,50138, HD\,45677, CD$-$24\,5721, CD$-$49\,3441, MWC\,623, and HD\,85567, 
have been previously considered as FS\,CMa stars, whereas HD\,181615/6 and HD\,98922 are added to the 
sample by this work.  We argue that the circumstellar grains of these isolated B[e] stars, 
already evolved beyond the pre-main sequence phase, should be formed in situ.  
This is in contrast 
to Herbig stars, which inherit large grains from parental molecular clouds. 
It has been thought that HD\,98922, in particular, is a Herbig star because of its large infrared excess,
but we propose it being in a more evolved stage.  Because dust condenses out of
stellar mass loss in an inside-out manner, the dusty envelope is spatially confined, and
anisotropic mass flows, or anomalous optical properties of tiny grains, lead to the generally
low line-of-sight extinction toward these stars. 
}
{}


   \maketitle
%
\section{Introduction}

B-type stars embrace a diversified stellar class.  In addition to the main sequence population, 
which itself covers a wide range of stellar luminosities and masses, some B-type stars 
exhibit emission lines in the spectra, show rapid rotation, or have gaseous/dusty envelopes.  
Yet because B-type stars evolve rapidly, it is often challenging 
to investigate their evolutionary status.  B[e] stars, characterized by additional forbidden 
lines in the spectra, and large near-infrared excess, are particularly elusive.  It is believed 
that their peculiarities originate from latitude-dependent mass loss with dusty disk-like 
structure \citep{zic03}.  

The B[e] phenomena are observed in heterogeneous stages of stellar evolution, from 
pre-main sequence (PMS) Herbig stars to evolved supergiants, compact planetary nebula or 
symbiotic stars \citep{lam98}.  Apart from these B[e] stars with established 
evolutionary status, there is a subclass still not well known, the 
"unclassified B[e] stars" \citep{lam98}, which have dust in their envelopes, yet lack cold dust components \citep{zic89}.  \citet{she00} and \citet{mir07} proposed that the 
unclassified B[e] stars with little cold dust, collectively called FS~CMa stars, 
are binary systems undergoing mass transfer.  The evolutionary stage of a star 
is relevant when its circumstellar dust is studied. 
In Herbig stars, the grains have progressively grown in size since in molecular clouds 
("old dust around a young star").  In contrast, the dust condensed out of the expanding
envelope of an evolved star must start with, and grow from, tiny grains  
("fresh dust around an old star").  This paper aims to address the evolutionary status 
of some of the unclassified B[e] stars.

Diagnostics of stellar maturity may not be always conclusive for single massive stars.  
Occasionally a star is presumed young on the basis of its large infrared excess alone.  
While such excess emission manifests existence of dust, it is a necessary, but not sufficient 
condition for stellar infancy.  Spectroscopic detection of enriched elements by 
nuclear synthesis in advanced stages of stellar evolution may differentiate a post-main 
sequence star from being in an earlier phase.  For example, the $^{13}$CO bandhead emission 
seen in the infrared K~Band has been used as indicative of an evolved status for 
a B[e] star \citep{kra09}.  Even though a few FS~CMa stars show the $^{13}CO$ feature, 
hence should be relatively evolved, some have evolutionary statuses that remain ambiguous.  


We started with a sample of nine Be stars with near-infrared excess rivaling to that of 
Herbig stars.  All our targets exhibit forbidden line spectra, therefore are B[e]-type stars, 
though in the literature, they might have also been classified as 
FS~CMa or Herbig stars.  In Section~\ref{sec:ire} we revisit the issue that 
thermalized dust in the circumstellar envelopes, instead of plasma free-free 
radiation, is responsible for the elevated infrared excess emission in Be stars.
In Section~\ref{sec:nocdust} we show the spectral energy distributions (SEDs) of our targets 
extending from mid-infrared to millimeter wavelengths to contrast the compact 
envelope sizes with those seen in typical Herbig stars.  
In Section~\ref{sec:belire} we present an elaborative census of each target in order 
to rule out the possibility of stellar youth for eight B[e] stars, adding support to 
their evolved status.  In Section~\ref{sec:fscma} we discuss the 
implication of evolutionary status and dust-formation mechanism in isolated 
B[e]-type stars. 


\section{Selection of targets}\label{sec:ire}

\begin{figure}
\begin{center}
   \includegraphics[width=\hsize]{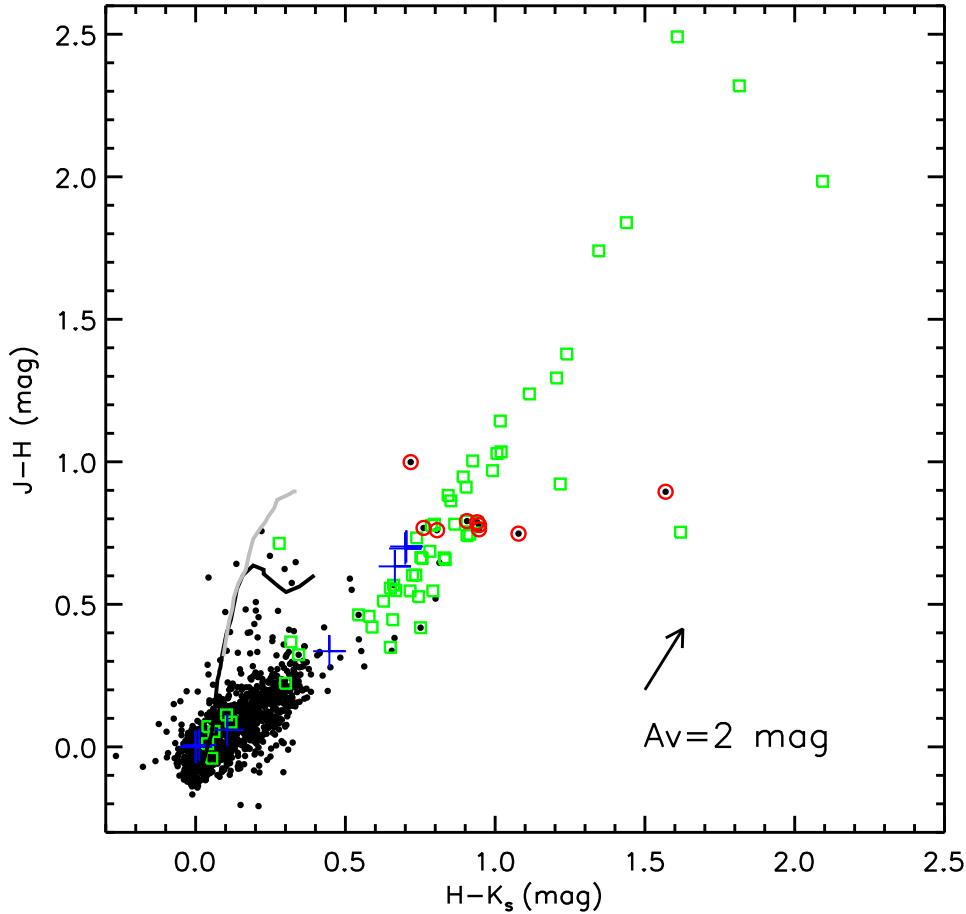}
 \caption{
     2MASS $J-H$ versus $H-K_{\rm s}$ diagram showing most Be stars (black dots) distinctly 
        separated from Herbig stars (squares), except a few Be stars with very large near-infrared 
        excess (open circles).  
        The sample of Be stars is taken from \citet{zha05}, and that of Herbig stars is from \citet{de01}.  
        The crosses indicate the levels of the free-free emission with varying electron number densities 
        and emitting volumes.  From bottom left, each cross represents an increase of an order, 
        from $10^0$ to $10^8$ times in emissivity (see text) surrounding a B0\,V star with a gas temperature of
        20,000~K. The first three crosses ($10^0$ to $10^2$ times) and the last two crosses ($10^7$ to $10^8$ times), 
        are, respectively, almost overlapped.  The gray and black curves are giant and dwarf loci 
       \citep{bes88} converted to the 2MASS system.      
       The arrow shows the reddening vector \citep{rie85} for an average Galactic total-to-selective 
       extinction ($R=3.1$).
                }
   \label{benire}
\end{center}
\end{figure}

Be stars typically have only moderate infrared excess arising from free-free
emission, with the majority being slightly redder than early-type main sequence stars, 
$J-H\lesssim 0.4$~mag and $H-K_{\rm s} \lesssim 0.4$~mag \citep{zha05,yu15}, as shown in 
Figure~\ref{benire}.  Some Be stars, however, exhibit near-infrared excess as large 
as that of Herbig stars, ${\rm H}-{\rm K} \gtrsim 0.7$~mag, so must contain dust in 
their envelopes \citep{all73}.  

Figure~\ref{benire} also shows the effect on 
near-infrared colors by adding free-free emission to a B0\,V photosphere, assuming an 
envelope gas temperature $T_g = 20,000$.  The envelope size together with electron number density 
alters the total emerging infrared flux, and hence also the $J-H$ and $H-K_{\rm s}$ colors. 
Adopting typical values of an emitting radius $R_0=10^{12}$~cm and electron density 
$n_0=10^{11}$~cm$^{-3}$ \citep{geh74,net82}, each cross in the figure represents the 
free-free intensity in orders from typical $n_{0}^{2} R_{0}^{3}$ (bottom left) to 
$10^{8}\, n_{0}^{2} R_{0}^{3}$ (top right), a value unreasonably large for 
a Be star.  The corresponding $J-H$ and $H-K_{\rm s}$ colors change from $\approx0$~mag, 
to an asymptotic value of $\sim 0.7$~mag, consistent with the results by \citet{all73}.  

Therefore, we selected among the compilation by \citet{zha05} Be stars with 
$J-H$ and $H-K_{\rm s}$ both $> 0.7$.  These include 
HD\,45677, HD\,259431, HD\,50138, CD$-$24\,5721 and CD$-$49\,3441, HD\,85567, HD\,98922, 
HD\,181615 and MWC\,623.  Note that except HD\,181615, all have been listed as 
Herbig stars \citet{the94}.  Moreover, other than HD\,181615 and HD\,98922, the other 
six are considered FS\,CMa stars \citep{mir07}.  Even though not selected 
by the forbidden line feature, our targets all turn out to be B[e] stars 
\citep{mir07,ack05,fin85,kip12}.  Our sample thus was chosen with no a priori assumption 
of the evolutionary status.  Table~\ref{tab:targets} summarizes the stellar parameters 
of our targets, collected from the literature, including the star name, coordinates, spectral 
type, reddening, and 2MASS J, H, and K$_s$ magnitudes.  


\begin{table*}
\smallskip
\centering
\large
\caption{Parameters of target B[e] stars}
\begin{tabular}{cccccccccc}\hline \hline
Star       &  R.A. (J2000)   & Decl. (J2000)               & SpType & $E(B-V)$ & J$^{a}$ & H$^{a}$ & K$_s^{a}$ \\
name       &  h~~~m~~~s      & $\degr~~~\arcmin~~~\arcsec$ &        &  (mag)   &   (mag) &   (mag) &    (mag)                                         \\
\hline\noalign{\smallskip}
HD\,45677     & 06 28 17.42 & $-$13 03 11.1 & B2$^{b}$                  & 0.2$^{c}$  & 7.24 & 6.35 & 4.78\\
HD\,259431    & 06 33 05.20 &    10 19 20.0 & O9$^{d}$                  & 0.42$^{e}$ & 7.45 & 6.67 & 5.73   \\
HD\,50138     & 06 51 33.40 & $-$06 57 59.5 & B7\,III$^{f}$             & 0.12$^{e}$ & 5.86 & 5.09 & 4.15 \\
CD$-$24\,5721 & 07 39 06.17 & $-$24 45 04.9 & B1.5$^{g}$                & 0.73$^{b}$ & 9.24 & 8.49 & 7.42 \\
CD$-$49\,3441 & 08 14 20.50 & $-$50 09 46.5 & B7$^{h}$                  & 0.3$^{h}$  & 8.98 & 8.22 & 7.41 \\
HD\,85567     & 09 50 28.53 & $-$60 58 03.0 & B2$^{h}$                  & 0.73$^{h}$ & 7.47 & 6.68 & 5.77  \\
HD\,98922     & 11 22 31.67 & $-$53 22 11.4 & B9V$^{i}$                 & 0.14$^{j}$ & 6.00 & 5.23 & 4.28 \\
HD\,181615/6  & 19 21 43.63 & $-$15 57 18.2 & O9\,V$+$B8p\,I$^{k}$      & 0.2$^{l}$  & 4.15 & 3.39 & 2.62  \\
MWC\,623      & 19 56 31.55 &    31 06 20.2 & B4\,III$+$K4\,I/II$^{m}$  & 1.4$^{c}$  & 7.10 & 6.10 & 5.38 \\
\noalign{\smallskip}\hline
\hline
\end{tabular}
\tablebib{
$^{(a)}$ 2MASS \citet{skr06}; $^{(b)}$ \citet{cid01}; 
$^{c}$ \citet{she00}; $^{(d)}$ \citet{vor85};
$^{(e)}$ \citet{fri92}; $^{(f)}$ \citet{ell15}; $^{(g)}$ \citet{mir03}; 
$^{(h)}$ \citet{mir01}; $^{(i)}$ \citet{car15}; $^{(j)}$ \citet{mal98}; 
$^{(k)}$ \citet{mal06}; $^{(l)}$~\citet{dyc72}; $^{(m)}$ \citet{zic01}
        }
\label{tab:targets}
\end{table*}

\section{Spectral energy distributions}\label{sec:nocdust}

\begin{figure}
\begin{center}
   \includegraphics[width=\hsize]{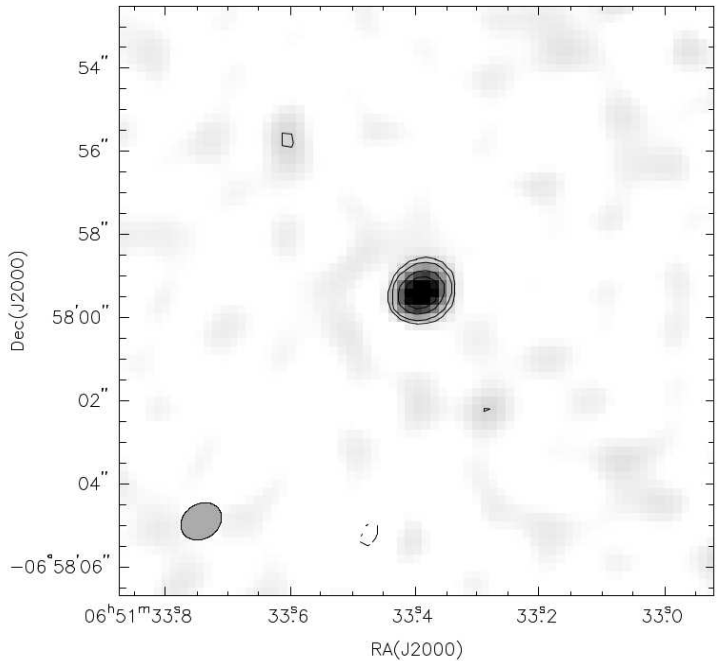}
 \vspace*{-0.5cm}
 \caption{SMA image of HD\,50138 at 225.8~GHz continuum with a 1$\sigma$ rms 0.9~mjy~beam$^{-1}$. 
        The contours are at $-3$, 3, 5, 10, 20$\sigma$ of the sky noise.  
        The lower left inset shows the synthesized beam.
              }
   \label{sma}
 \end{center}
\end{figure}

The SEDs of our targets have all been presented in the 
literature \citep{de97,mal98,mir03,mir01,zic89,net09,whe13,kra08b,kra08a,ell15}. At long wavelengths,  
HD\,50138 lacked data, so we observed it at 225.8~GHz with the 
sub-millimeter array (SMA) on February 13, 2012.  The observations were carried out 
with an extended configuration of baselines ranging 
from 44~meters to 225 meters.  The radio sources 0530$+$135 and 0607$-$085 were used as antenna 
complex gain calibrators, whereas 3C\,84 and Callisto were employed for bandpass and flux calibration, 
respectively.  The typical uncertainty in the absolute flux scale is 20\%. With an on-source integration
time of 1.25~hours, a total bandwdith of 8~GHz in the double-side-band mode provided by the digital 
correlator for continuum imaging, and typical system temperatures ranging between 160~K and 250~K, 
an rms noise of about 0.9~mJy/beam was reached. Figure~\ref{sma} shows the 1.3~mm continuum image 
of HD\,50138, with the synthesized beam of $1\farcs03  \times 1\farcs03$ obtained with natural weighting. 
The source was unresolved with a peak flux of 20.75~mJy/beam, yielding an S/N greater than 20.


\begin{figure*}[t!] 
     \resizebox{0.33\hsize}{!}{\includegraphics{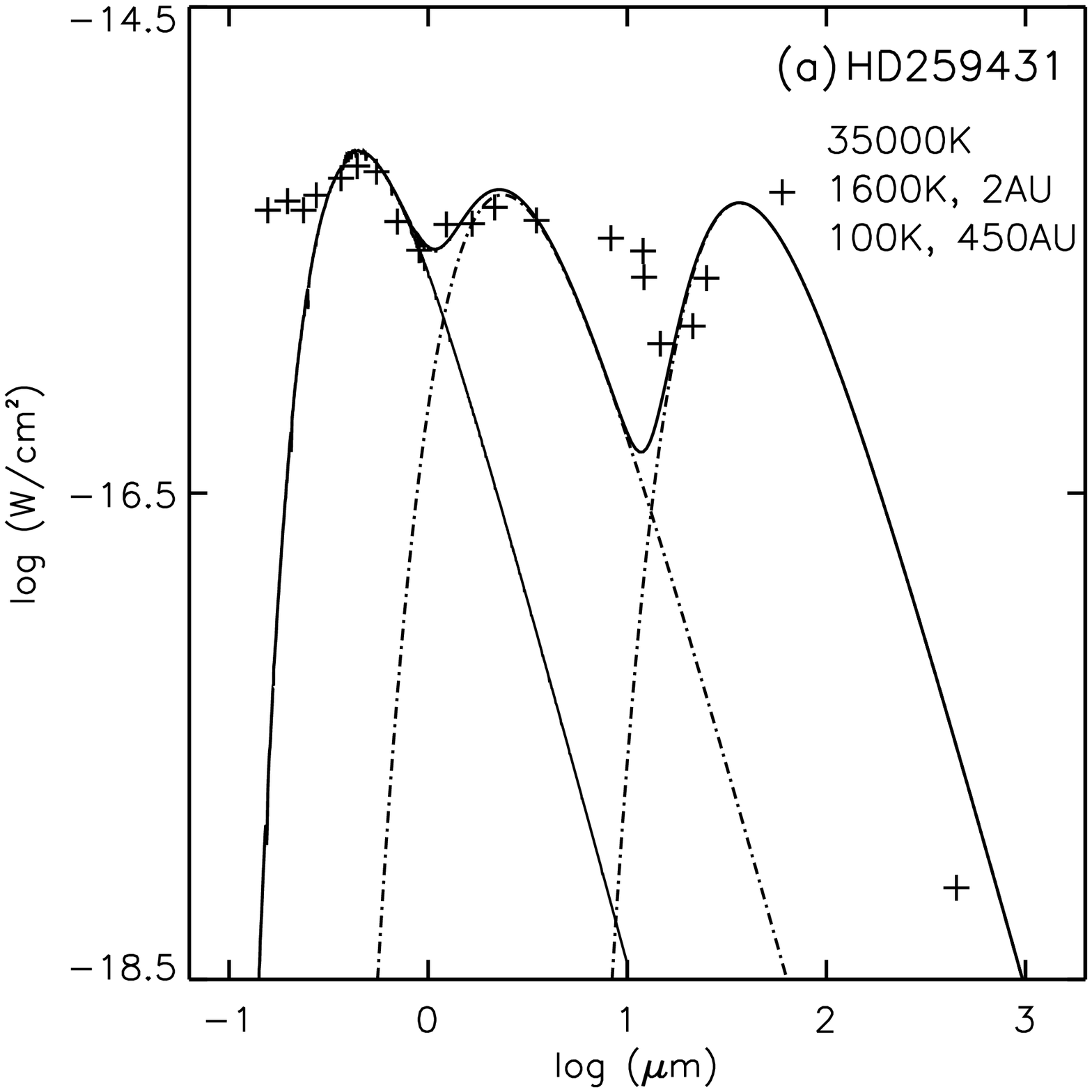}}
  \resizebox{0.33\hsize}{!}{\includegraphics{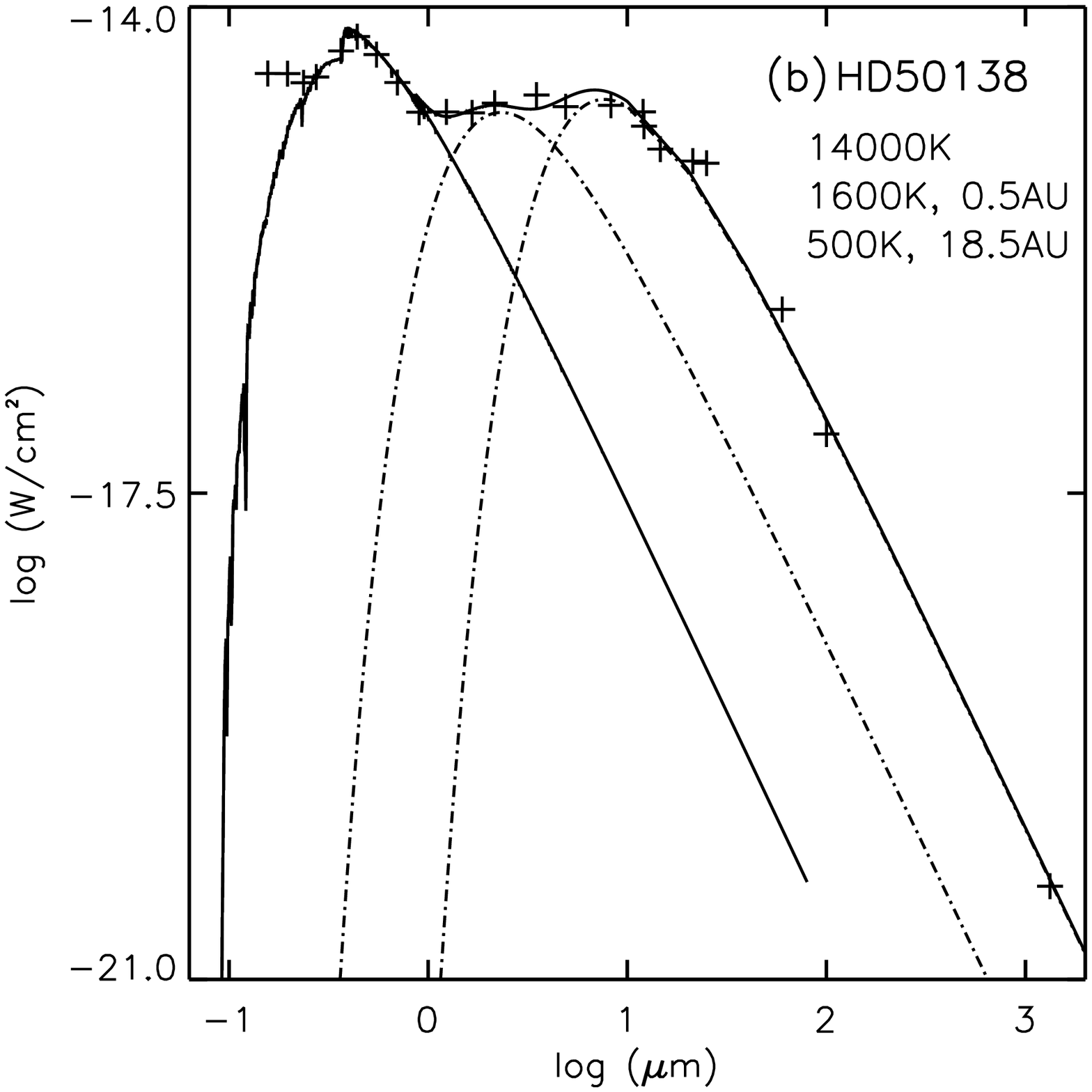}}
 \resizebox{0.33\hsize}{!}{\includegraphics{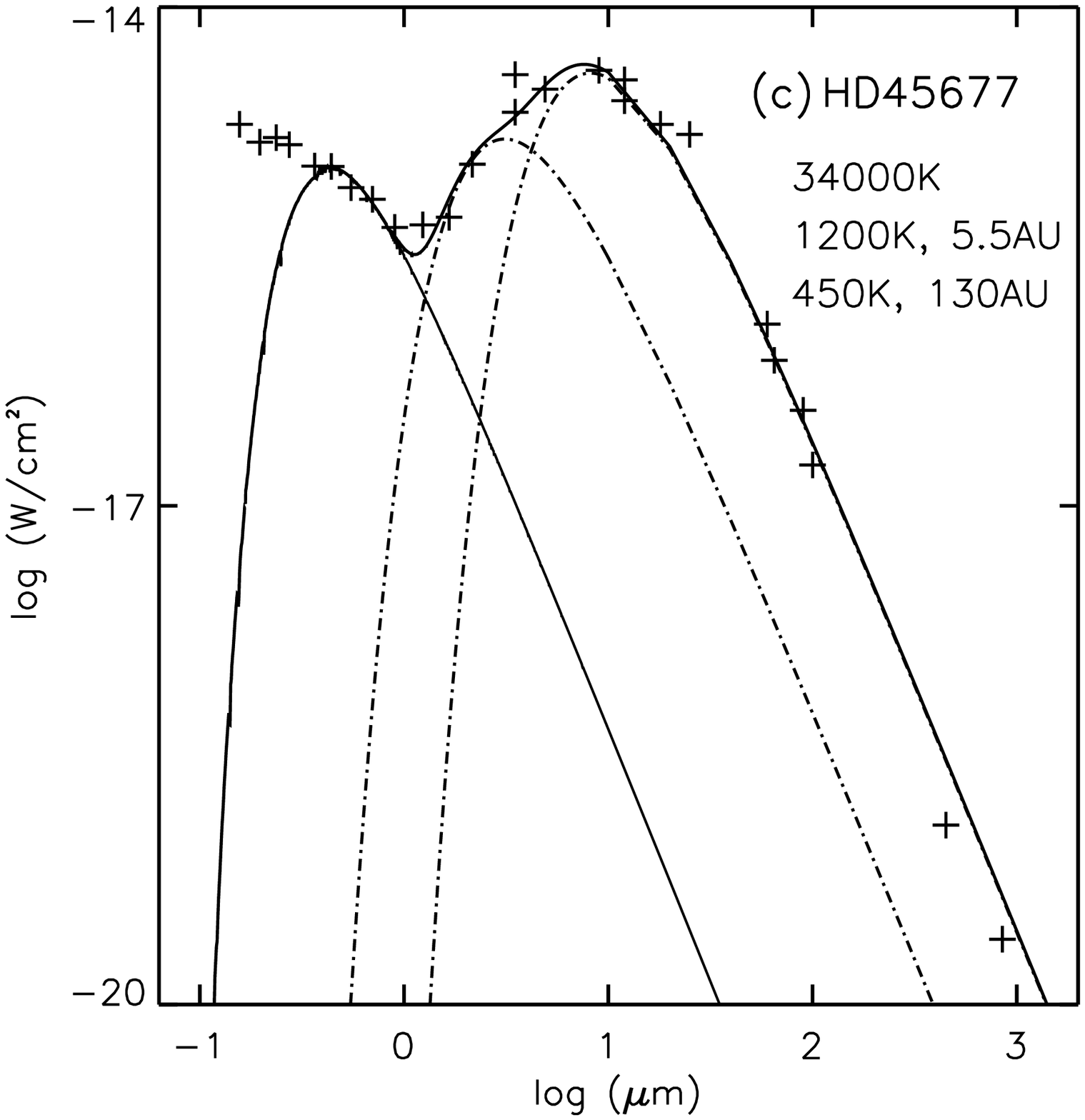}} 
  \resizebox{0.33\hsize}{!}{\includegraphics{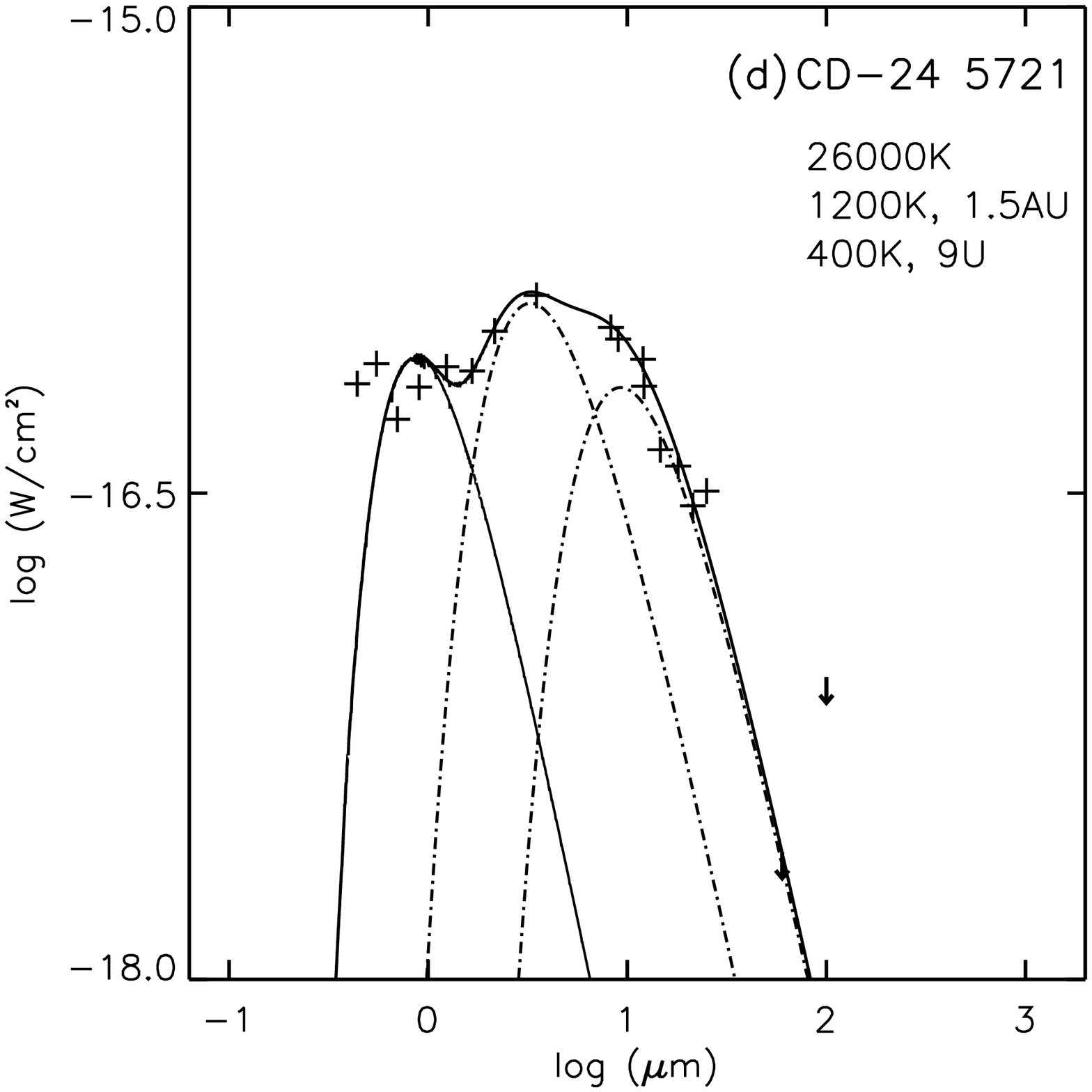}}
    \resizebox{0.33\hsize}{!}{\includegraphics{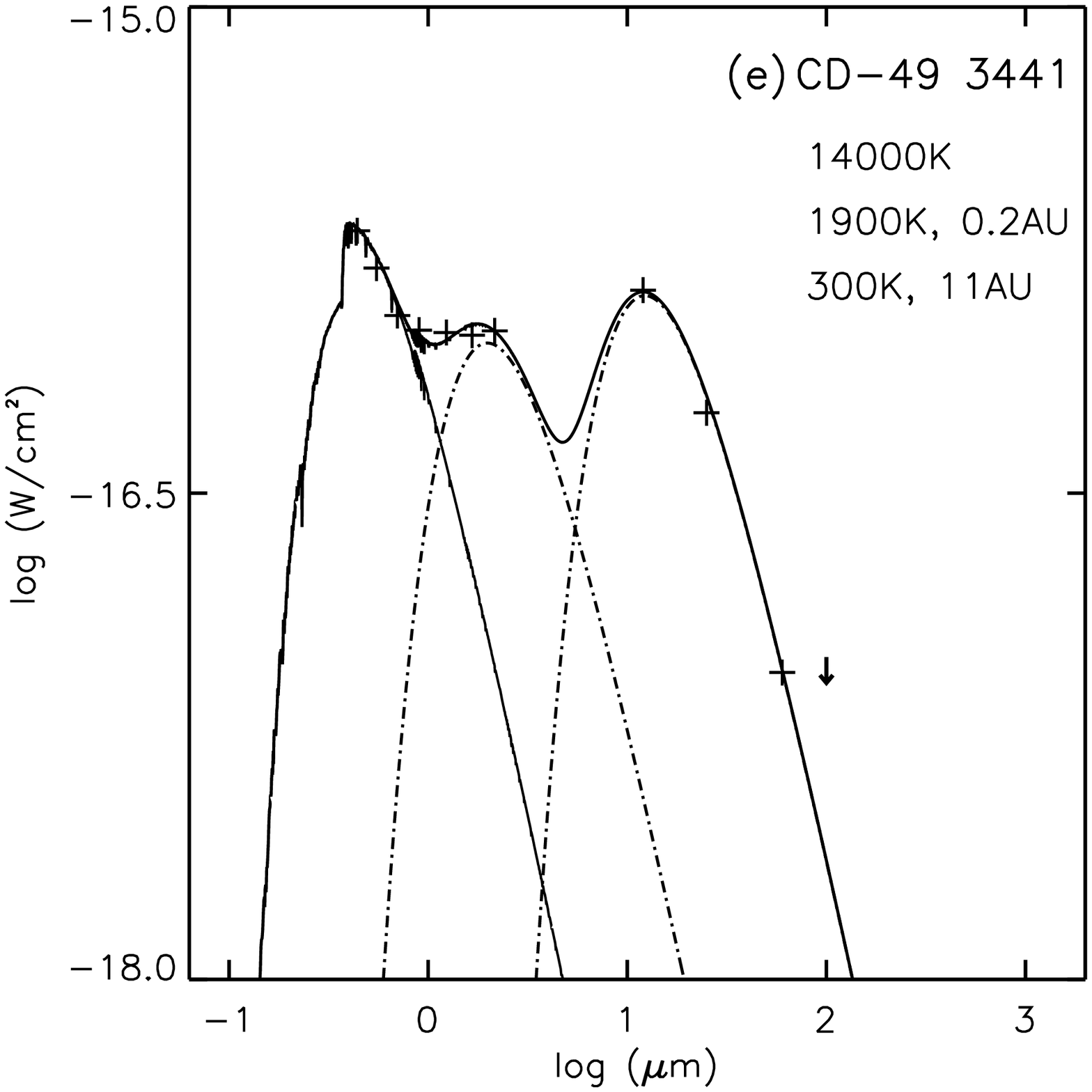}}
    \resizebox{0.33\hsize}{!}{\includegraphics{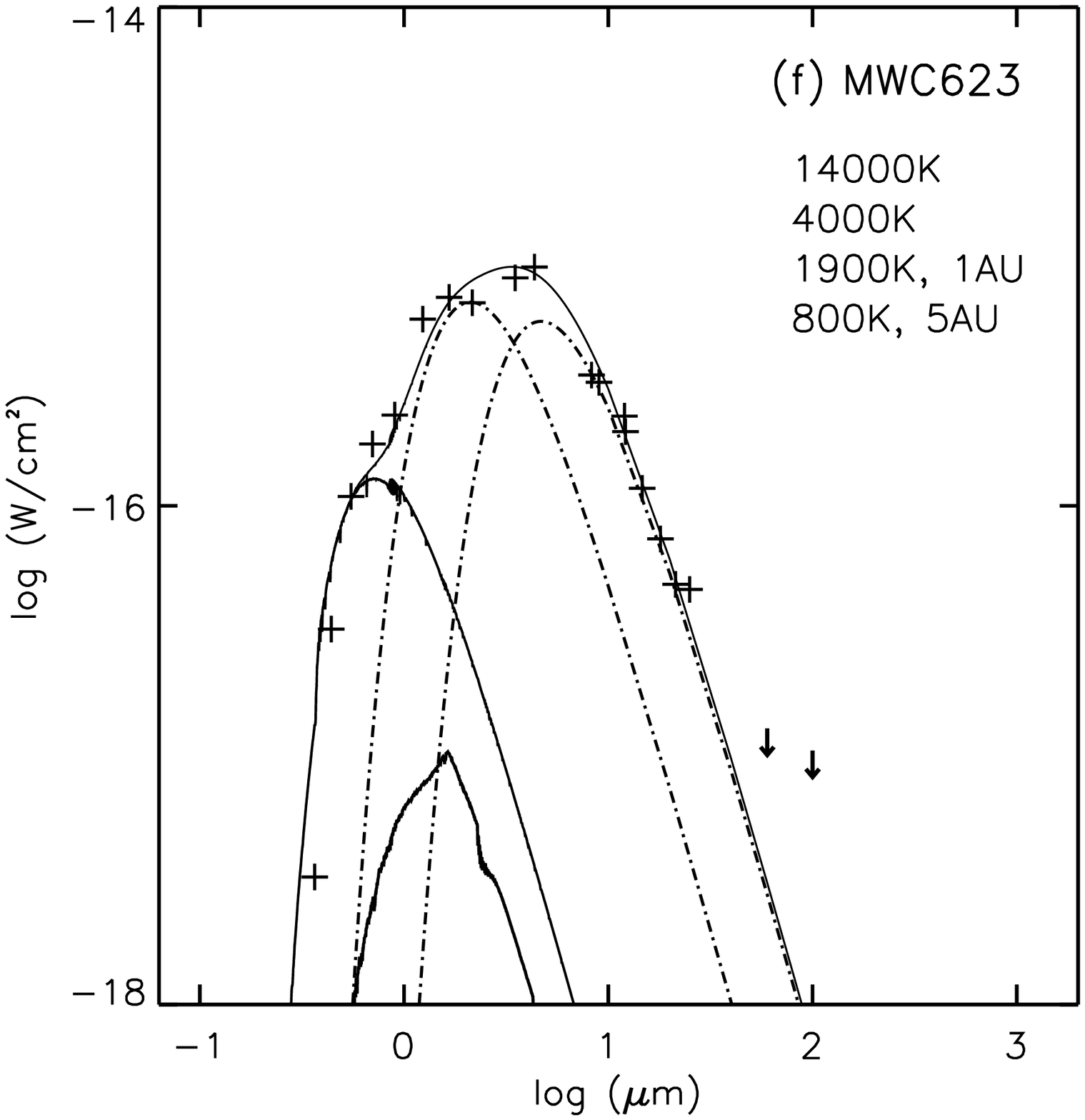}}\\
     \resizebox{0.33\hsize}{!}{\includegraphics{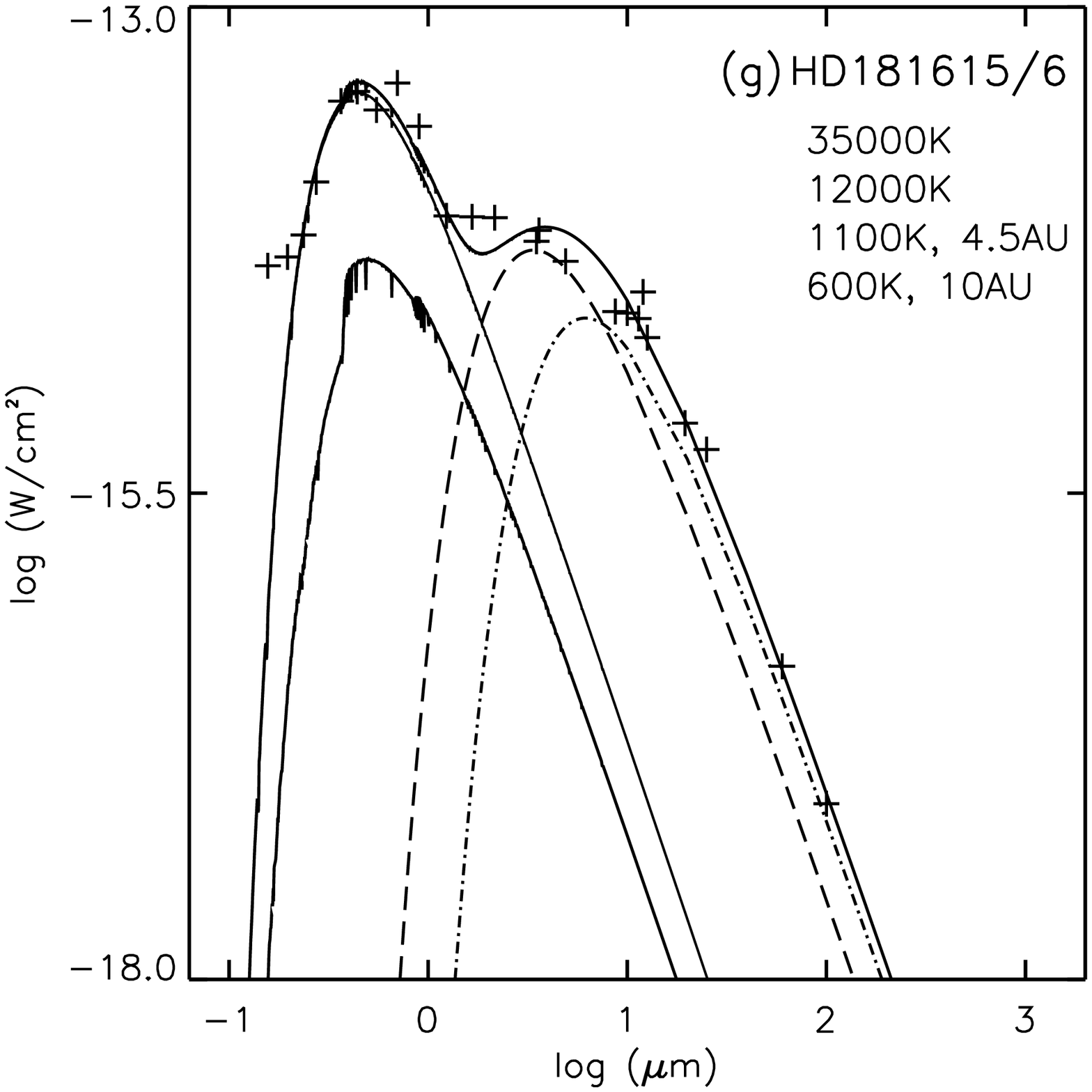}}
  \resizebox{0.33\hsize}{!}{\includegraphics{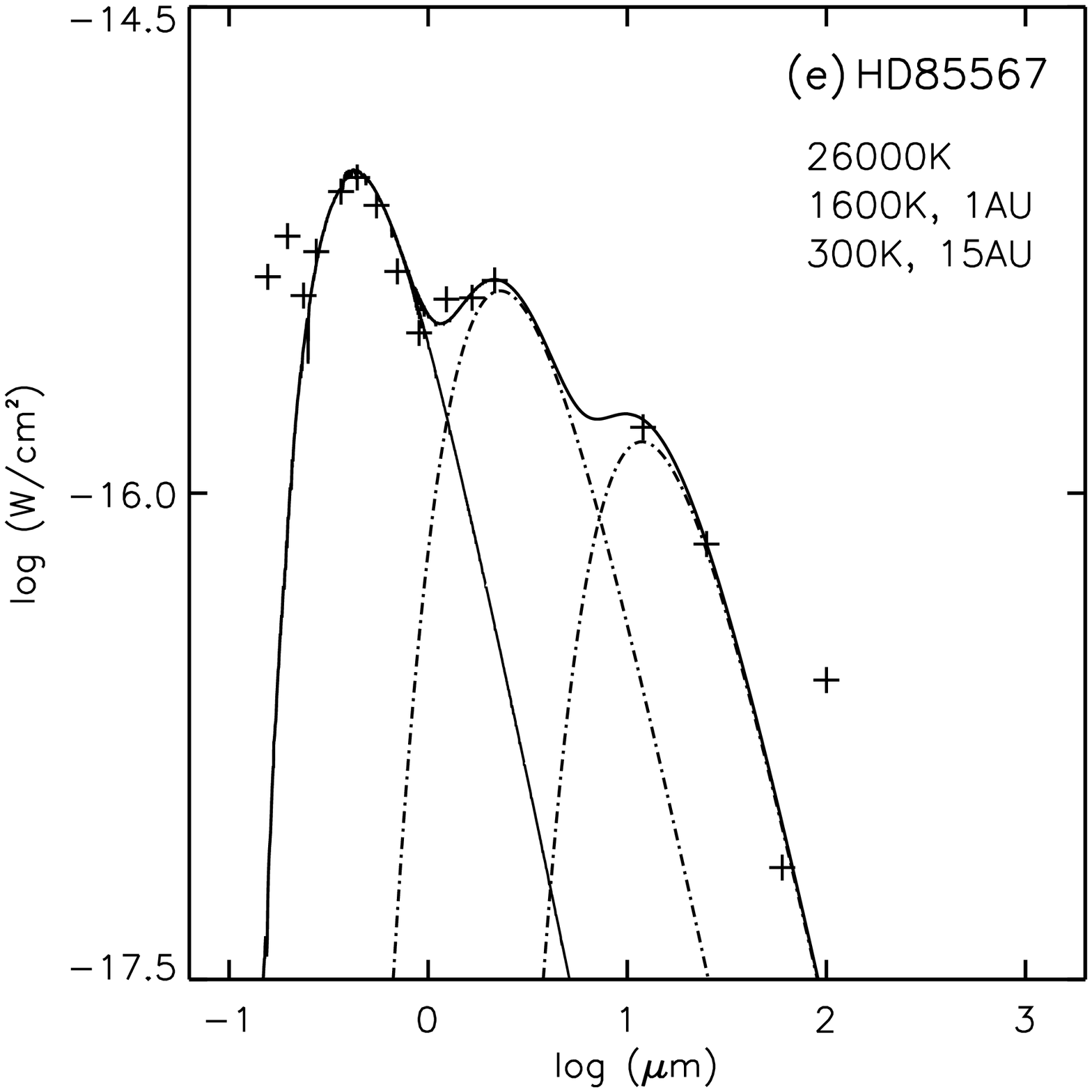}}
     \resizebox{0.33\hsize}{!}{\includegraphics{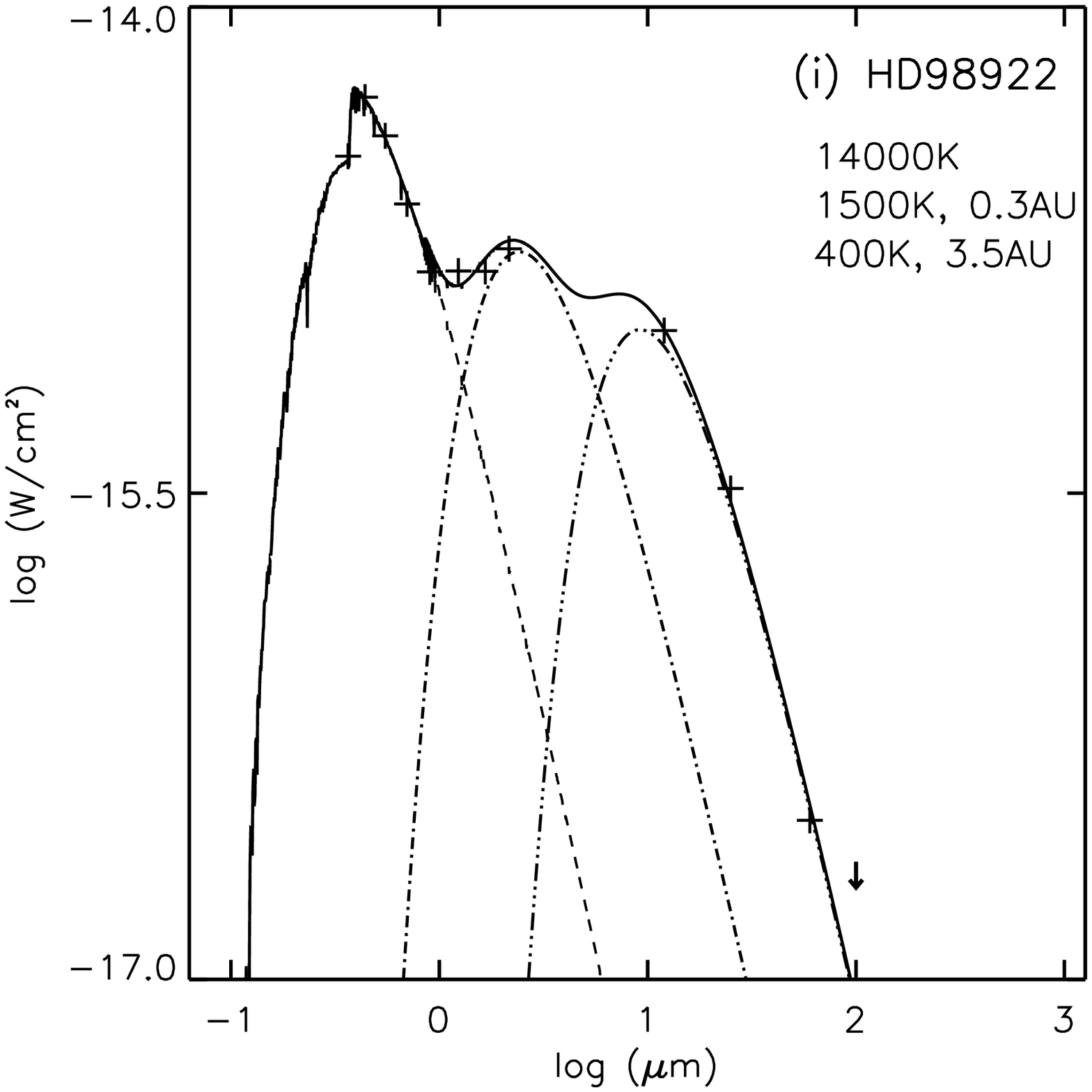}}
 \caption{
    SEDs of B[e] stars showing large near-infrared excess emission.
    (a) The Herbig star HD\,259431 has a prominent excess emission not only in near-infrared, but 
    also in far-infrared wavelengths.  (b)--(i) The sample of B[e] stars with reduced excess emission 
    in far-infrared and submillimeter.  Overlaid for each star are a model stellar photosphere 
    \citep{kur93} (solid line), two blackbody curves to present the inner and outer edges of the 
    envelope (dash-dotted lines), and the sum of the three (top solid line).  Photometric data sources 
    are summarized in Tables~\ref{tab:targets} and \ref{photo}.  Additional data include those in 
    (a)~\citet[][450~$\mu$m]{man94}, (b)~1.3~mm measurement (this work), in 
    (c)~AKARI/IRC mid-infrared all-sky survey and AKARI/FIS all-sky survey point source catalogs 
    \citep[][9, 18, 65 and 90~$\mu$m]{yam10,Ish10}, COBE DIRBE point source catalog 
    \citep[][3.5, 4.9 and 12~$\mu$m]{smi04} and \citet[][450 and 850~$\mu$m]{di08}, 
    and in (g)~\citet[][2.3, 3.6, 4.9, 8.7, 10, 11.4, 12.6 and 19.5~$\mu$m]{geh74}. 
    }
 \label{sed}
\end{figure*}

Figure~\ref{sed} shows the SEDs of our targets, each typified with a near-infrared excess, followed 
by a steep decrease toward mid- and far-infrared wavelengths.  The only exception is HD\,259431, 
which shows prominent excess emission extending to millimeter wavelengths.  For each target,  
an atmospheric model \citep{kur93} reddened by the observed $Av$ value \citep{zha05, nec80}, 
was used to approximate the stellar photosphere with its adopted spectral class.  
Two additional components, each of a blackbody radiation, were added to represent the inner and 
outer edges of the envelope, to approximate each SED at long wavelengths.

\begin{sidewaystable*}
\normalsize
\smallskip
\caption{Photometric data}\label{photo}
\begin{tabular}{lrrrrrrrrrrrrrr}
\hline\hline\noalign{\smallskip}             
Star & F1565$^{a}$ & F1965$^{a}$ & F2365$^{a}$ & F2740$^{a}$ 
& U & B & V & R & I & L$^{b}$ & [12]$^{c}$ & [25]$^{c}$ & [60]$^{c}$ & [100]$^{c}$ \\
name & (W/m$^{2}$/nm) & (W/m$^{2}$/nm) & (W/m$^{2}$/nm) & (W/m$^{2}$/nm) 
& (mag) & (mag) & (mag) & (mag) & (mag) & (mag) & (Jy) & (Jy) & (Jy) & (Jy) \\
\noalign{\smallskip}\hline\noalign{\smallskip}
HD\,45677     &    1.26E$-$13    &    7.84E$-$14    &    6.94E$-$14    &    5.43E$-$14    &    7.86$^{b}$    &    8.52$^{b}$    &    
8.5$^{b}$    &    8.11$^{d}$    &    8.01$^{d}$    &    2.15    &    146.00    &    143.00    &    24.80    &    5.87    \\
HD\,259431    &    2.95E$-$14    &    2.56E$-$14    &    1.95E$-$14    &    1.94E$-$14    &    8.48$^{e}$    &    9.01$^{e}$    &    
8.73$^{e}$    &    8.72$^{d}$    &    8.61$^{d}$    &    4.58    &    12.54    &    20.19    &    109.40    &    1.13    \\
HD\,50138    &    2.12E$-$13    &    1.69E$-$13    &    1.20E$-$13    &    1.14E$-$13    &    6.26$^{f}$    &    6.66$^{f}$    &    
6.64$^{f}$    &    6.58$^{d}$    &    6.59$^{d}$    &    2.72    &    70.30    &    62.50    &    13.30    &    2.81    \\
CD$-$24\,5721    &        &        &        &        &        &    11.53$^{g}$    &    11.04$^{g}$    &    10.90$^{h}$    &    
10.13$^{h}$    &    5.86    &    3.28    &    2.67    &    0.49    &    2.86    \\
CD$-$49\,3441    &        &        &        &        &        &    10.35$^{d}$    &    10.3$^{g}$    &    10.30$^{h}$    &    
10.31$^{h}$    &        &    5.35    &    4.67    &    1.77    &    3.29    \\
HD\,85567    &    2.97E$-$14    &    3.16E$-$14    &    1.72E$-$14    &    2.03E$-$14    &    8.14$^{f}$    &    8.69$^{f}$    &    
8.57$^{f}$    &    8.51$^{d}$    &    8.46$^{d}$    &        &    6.39    &    5.81    &    1.40    &    8.84    \\
HD\,98922    &        &        &        &        &        &    6.62$^{i}$    &    6.82$^{i}$    &    6.78$^{d}$    &    
6.74$^{d}$    &        &    40.16    &    27.24    &    6.19    &    7.69    \\
HD\,181615/6    &    2.98E$-$13    &    4.10E$-$13    &    3.60E$-$13    &    2.60E$-$13    &    4.18$^{f}$    &    4.71$^{f}$    &    
4.61$^{f}$    &    4.54$^{d}$    &    4.52$^{d}$    &        &    137.00    &    44.20    &    8.15    &    2.67    \\
MWC\,623    &        &        &        &        &        &    11.99$^{j}$    &    10.5$^{j}$    &    10.34$^{d}$    &    
9.79$^{d}$    &    3.85    &    9.16    &    3.85    &    2.56    &    3.48    \\
\noalign{\smallskip} \hline
\hline
\end{tabular}
\tablebib{
$^{(a)}$ \cite{tho78}; $^{(b)}$ \cite{mor78}; $^{(c)}$ \cite{hel88}; 
$^{(d)}$ \cite{mon03}; $^{(e)}$ \cite{her82};  $^{(f)}$ \cite{mer87a,mer87b}; 
$^{(g)}$ \cite{amm06}; $^{(h)}$ \cite{den05}; $^{(i)}$ \cite{mye01}; 
$^{(j)}$ \cite{per97}
}
\end{sidewaystable*}


%

\section{Census of neighboring regions}\label{sec:belire}


Because a Herbig star, unlike a T~Tauri star, lacks unambiguous youth discriminant such as the 
lithium absorption in the spectrum, it is difficult to distinguish a PMS 
Herbig star from an 
early-type main sequence star.  Nonetheless, the short PMS lifetime, for example, 10~Myr at most 
for a B-type star \citep{lej01}, means a runaway Herbig star cannot transverse far from its parental 
cloud.  While an early-type star associated with star formation can already be on the main sequence, 
one that is isolated from any nebulosity, dark clouds, or young stellar population must have evolved 
beyond the PMS phase.   Therefore, we conducted a thorough neighborhood
census of our targets in order to resolve their evolutionary status, in response to the question raised by \citet{her94} 
 ``are there stars which look like Ae/Be's yet are not pre-main-sequence?'' 

{\textbf HD\,259431:} This star, with a known companion $\sim3\arcsec$ away \citep{bai06}, is seen near the 
center of the reflection nebula NGC\,2247, with many nearby star formation regions (SFRs) 
projected within a few deg, 
for example, NGC\,2245, DOG\,105, IC\,2167, LDN\,1607, and NGC\,2264.  The spectral type reported in the 
literature ranges from O9 to B6 \citep{vor85,fin85,her04}, so its PMS phase, in any case, must be short.  
Its SED shows copious excess emission leveling to far-infrared, the most prominent, and hence with the most 
extended outer dust envelope, among our targets.  The star has 
a Hipparcos distance of 173~pc \citep{van07}.  The cluster NGC\,2247 has a distance 800~pc, estimated by the 
CO line radial velocity, similar to that of Mon\,OB1 projected 2~deg away \citep{oli96}.  
HD\,259431 therefore should be in the foreground of NGC\,2247.  However, given the uncertain distance determination 
of, and the line-of-sight alignment with, NGC\,2247, we cannot rule out the possible association of the star  
with the nebulosity.  HD\,259431 may well be a PMS Herbig \citep{ale13} or a young main sequence 
star, so is excluded in our final sample of relatively evolved B[e] stars.

We emphasize that our targets were selected on the basis of the level of infrared excess.  The Be star sample by 
\citet{zha05} precludes Herbig stars known at the time, but is not conclusive.  That is, Herbig stars 
may still be included, and some FS\,CMa stars historically recognized as Herbig stars need further 
scrutiny.  Except HD\,259431 just presented, the other eight targets turn out to be isolated B[e] 
stars.  Their neighborhood census follows.

{\textbf HD\,50138:} Also known as MWC\,158, this is a B7\,III star \citep{ell15} at a distance 392~pc 
measured by Hipparcos \citep{van07}.  
There are some distant SFRs, projected about one degree away, including the \mbox{H\,{\sc ii}} 
regions LBN\,1040 and BSF\,60, the reflection nebula GN\,06.47.6.01, and the T\,Tauri
star J06495854$-$0738522, all at distances more than 1~kpc \citep{qui06,fic84,mag03,vie03}.  
Additional SFRs with no distance information include Paramian\,16, Gn\,06.54.8.01, Gn\,06.54.8.03, 
and 220.9$-$2.5A.  There are four open clusters, M\,50, NGC\,2302, NGC\,2306 and NGC\,2309, 
within 3~deg and with ages 53~Myr \citep{lok01}, 309~Myr \citep{kha05}, 800~Myr \citep{kha05} 
and 250~Myr \citep{pia10}, respectively.  Among these, NGC\,2302 is only 8~arcmin off 
with a heliocentric distance about 1200~pc \citep{kha05}, whereas other clusters are all beyond 
4.5~deg away and at a distance 1150~pc.  The spatially extended CMa\,OB1 \citep[age 3~Myr,][]{cla74} 
is seen about 4.5~deg away and at a distance 1150~pc.  
We cannot rule out HD\,50138 being a possible 
main sequence
member escaped from one of 
these clusters, given the main sequence lifetime of some five hundred~Myr with its spectral type.    
The star has a measured rotational speed $90~km~s^{-1}$, and 
is inferred to have a disk inclined at $56\degr$ \citep{bor11,bor12}.
Using the spectrointerferometric technique, \citet{ell15} 
resolved its au-scale disk with a measured photocenter offset, possibly caused by an asymmetric 
structure of the disk or a binary companion.  These authors favored a post-main sequence status
for HD\,50138, because of the presence of high transition lines, such as Paschen, Bracket 
and Pfund series less commonly observed in PMS stars, and because the star is apparently 
not associated with an SFR, supported by our assessment. 


{\textbf HD\,45677:} The spectral type reported for this star ranges from B0, B2\,IV/V[e] to B3 
\citep{tuc83,cid01,boh94}.  At a distance 279~pc \citep{van07}, this star has the largest near-infrared 
excess among our targets.  \citet{zic03} inferred that HD\,45677 is nearly pole-on based on the high resolution 
spectroscopic line profiles.  Its infrared spectra indicate existence of not only silicate grains \citep{mol02a,mol02b}, 
but maybe also calcium silicate hydrate, that is, cementitious nano-particles, as suggested by modeling the 
10 and 18~$\mu$m features by \citet{bil14}, rendering evidence of ultra-small particles in the envelope.  

\citet{her94} described this star as ``projected upon completely unobscured fields''.  
The nearest SFR WV\,340 is projected some $45\arcmin$ away but with no distance information.   
Two other SFRs, CMa\,OB1 and Mon\,R2, are eight 
deg away and at distances more than 800~pc \citep{mel09,her75}, so too far to be related to HD\,45677.  
Besides, both these SFRs are a few Myr old \citep{cla74, and06}, too young for such an early-type star 
to be an escaped member.  Hence, HD\,45677 should not be a PMS Herbig star.  The star, also called FS\,CMa, 
characterized by spectral forbidden lines and strong Balmer emission 
\citep[e.g.,][]{de97}, is the prototype of the subclass of Be stars studied by \citet{mir07}.  

{\textbf CD$-$24\,5721:} This B1.5\,V star is at a distance 3.5~kpc \citep{mir03}.   
\citet{zic03} reported narrow \mbox{Fe\,{\sc ii}} absorption lines resembling classical Be stars, 
yet claimed the star as being an unclassified B[e] star because of the presence of P~Cygni profiles 
in Balmer lines and \mbox{He\,{\sc i}} line variability.  
There are only three small clouds, LDN\,1668, TGU\,H1615 and 
240.25$-$1.75 \citep{may97}, plus one open cluster C\,0739$-$242 \citep{dri91, cid01}, seen 
within 1~deg, and a large ($\sim 0.4$~square degrees) dark cloud LDN\,1667 about 2.5~deg to the south-west.  
None of these have distance information in the literature.  The star is some 4~deg away from the 
large OB association, Pup-CMa \citep[520~pc,][]{gyu99} and is 1.5~deg away from the open cluster M\,93, 
which is 400~Myr old \citep{ham00} and at a distance of 1037~pc \citep{kha05}.  Another open cluster 
NGC\,2467, with an age 12.6~Myr \citep{hua06} and at a distance 1355~pc \citep{kha05}, is 3.25~deg away.  
Both these clusters should not be related to CD$-$24\,5721, so the star should not be young.


{\textbf CD$-$49\,3441:} Also known as Hen\,3$-$140, this star has an uncertain spectral type.  With high 
resolution spectroscopy, \citet{mir01} inferred a B7 type in terms of the intensity ratios of He lines, 
but a B2 type in terms of the Balmer line wings.  Based on radial velocity and interstellar extinction, 
the estimated distance is approximately 2~kpc \citep{mir01}.  Even with a late-B type, the star as a dwarf
would be at 1~kpc.  There are molecular cloud BHR\,31 and reflection nebula BRAN\,137, 
both about 1~deg away at $\sim400$~pc \citep{hen98,sta89}.  
The open cluster NGC\,2547, with an age 38.5~Myr and a distance 361~pc \citep{nay06}, is 1~deg away.   
All these objects are within the Gum nebula which itself extends about 120~pc in radius at an average 
distance 400~pc \citep{sus11}, so not related to CD$-$49\,3441.  


{\textbf MWC\,623:} This is a binary system at $\sim2$~kpc \citep{zic01} with a B4\,III star and 
a K4\,I/II companion \citep{lie14}.  The binary period is uncertain with long-term radial velocity monitoring 
\citep{zic01,pol12}, and only a minimum period of 25 years was suggested \citep{pol12}.  
Among our targets, only HD\,181615/6 and MWC\,623 are known with spectral types for both the primary and secondary. 
MWC\,623 is 1.26~deg above the Galactic plan, with eight unresolved sources within 20~arcmin \citep{tay96}, 
plus two young stellar objects and one supernova remnant projected within 20--40~arcmin.  None of these 
has a distance estimation except a very rough 7--17~kpc estimated for the supernova remnant \citep{hui09}, 
but their small sizes suggest likely being more distant than, thus unrelated to, MWC\,623.  

Additionally, the prominent \mbox{H\,{\sc ii}} region W\,58 and the young stellar object IRAS\,19592$+$3302, 
both more than 2~deg away, 
are associated with the Perseus arm or Cygnus arm, which are at 8.3~kpc and 13.38~kpc, respectively 
\citep{tho06,yan02}, so not related to MWC\,623.  In addition to SFRs, one old open cluster 
(about 800~Myr) Kronberger\,52 is $24\arcmin$ away, but is at 3.6~kpc \citep{sza10}, so too far 
for MWC\,623 to be an escaped member.  Accordingly, MWC\,623 should not be a PMS star.  The cool 
stellar companion should contribute to the near-infrared excess, but in no way could extend beyond mid-infrared.  
MWC\,623 therefore should be a secure case of a non-PMS dusty Be star. 
The observed extinction of MWC\,623, $E(B-V)=1.4$, the largest among our targets, is almost entirely  
interstellar \citep{she00}, suggesting little dust extinction along the line of sight to the star.


{\textbf HD\,181615/6:} Also known as $\upsilon$~Sgr, this star is an evolved spectroscopic binary 
\citep{sch83,cam99} consisting of a luminous supergiant and an ``invisible'' but hotter and more 
massive dwarf with a binary period 138~d \citep{wil15} and mass ratio $q=0.63 \pm 0.01$ \citep{dud90}.  
The spectral type reported for the supergiant primary ranges from B5\,II, A2\,Ia to F2\,I 
\citep{bon11,abt79,mau25}, and for the dwarf secondary from O9\,V to B\,3V \citep{mal06,mau97}.  
With a close distance of 595~pc \citep{van07}, the circumbinary geometry of HD\,181615 was well 
determined by the optical and mid-infrared interferometric techniques \citep{bon11,net09}:~the 
radius of the bright companion ($20~R_\odot$), the H$\alpha$ formation region (12~au), the 
inner radius of the geometrically thin disk (6~au), the angle of disk inclination (50\degr) and 
orientation (position angle $80\degr$).  
There are no obvious SFRs within 10 deg from the star, with only four small or unresolved radio sources,
OV$-$133, OV$-$136, PMN\,J1922$-$1525 and PMN\,J1924$-$1549 \citep{ehm70,gri94}, 
projected about 0.5~$\deg$ away.  None of the them have distance estimation in the literature.
There is little doubt that this is an evolved system, so the dust should be freshly produced in situ, 
likely as a result of mass transfer or wind-wind collision.  


{\textbf HD\,85567:} This B2 star is at a distance 1.5~kpc \citep{mir01}.  
\citet{whe13} found no indication of close binary within 100~AU using near-infrared spectrointerferometry.  
These authors found the inner radius of the dusty disk to be undersized ($\sim1$~au) given the 
stellar luminosity, and attributed it to the presence of dense gas in the interior of the disk.  
Moreover, they suggested the cold dust in the outer disk to have been photoevaporated so its 
SED shows a strong near-infrared excess but little far-infrared excess.  Our SED analysis 
indeed shows an inner envelope size less than 1~au with a temperature 1600~K, and an outer envelope edge 
of 15~au at 300~K.  

There are two high-velocity clouds \citep{wak91,put02} seen within 50\arcmin, but they are small in 
angular sizes so likely in the background and not physically associated with HD\,85567. 
The open cluster NGC\,3114, at a distance of 820~pc \citep{gon01} and with an age of 160~Myr, 
is some 1.3~deg away, is too old to harbor any Herbig stars.  Therefore, HD\,85567 is very unlikely to 
be in the PMS phase.

{\textbf HD\,98922:} This star is the most isolated among our targets, with only the compact 
dark nebula G291.11$+$7.86 and the reflect nebula BRAN\,350 projected 1.4~deg away, neither with distance
information.  The prominent SRF, the Carina Nebula, projected 7~deg away, is at a distance 2.7~kpc \citep{tap03}.

HD\,98922 was considered a Herbig star on the basis of its infrared excess \citet{the94}, which has been 
allegedly propagated through the literature, with no further evidence of stellar youth.  
\citet{hal14} detected tenuous but extended (more than 600~au across) 
molecular gas.  Little reddening to the star, $E(B-V)\sim0.07$--$0.14$ \citep{mal98,hal14},  
indicates non-spherical distribution of circumstellar dust.  It is puzzling how the star 
could form out of a completely isolated cloud, keep the surplus gas and dust, and 
maintain active accretion \citep{ale13}.  

We note that the distance determination, 400 or 500~pc, for neither a B9\,V nor an A2\,III spectral type 
\citep{car15,hal14} can reconcile with the {\it Hipparcos} measurement \citep[1150~pc][]{van07,ale13}.  
Using high-dispersion spectroscopy, \citet{hal14} derived a gravity ($\log=3.0$) and effective 
temperature ($T_e=9000$~K) that put the star much above even the 1~Myr isochrone in the HR diagram.  
This inconsistency can be vindicated if this star is in the post-main sequence phase.  
Adopting the effective temperature and stellar luminosity \citep{hal14} the star is indeed consistent 
with being a $\sim5$~M$_\odot$ star along the giant branch, likely a luminosity class II bright giant.  
In any case, HD\,98922 should not be a PMS object.

Apart from positional isolation, our targets should not be associated with any SFRs in terms of 
space motions.  Table~\ref{tab:pmrv} lists the heliocentric distances, proper motions, and radial 
velocities of our targets.  
The three stars having complete proper motion and radial velocity measurements have Galactic 
space motions ($U$, $V$, $W$), respectively of $(7.0, -1.8, 4.9)$ for HD\,45677, $(8.8, 5.3, 4.0)$
for HD\,259431, and $(23.8, -0.8, 2.6)$ for HD\,50138, all in km~s$^{-1}$, which are 
within the range for field disk-population stars.  

HD\,85567 and CD$-$49\,3441 are intrinsically high-velocity stars.  HD\,85567 has a relative 
fast $U\sim69$~km~s$^{-1}$.  Notably, CD$-$49\,3441, with its appreciable 
proper motion $\sim20$~mas~yr$^{-1}$ \citep{hog00} and a large distance yield a tangential space velocity 
close to 200~km~s$^{-1}$ alone.  Plus a radial velocity 33~km~s$^{-1}$ \citep{mir01}, it is 
therefore not impossible that CD$-$49\,3441 was escaped from one of the regions projected on
the sky or along the line of sight.


\begin{table*}
\smallskip
\centering
\large
\caption{
Distances, proper motions and radial velocities}
\begin{tabular}{cccccccccc}\hline \hline
Star       &  dist & $\mu_\alpha$       & $\mu_\delta$   & RV   & References           \\
name       &  (pc) & mas\,yr$^{-1}$     & mas\,yr$^{-1}$ & km~s$^{-1}$     &  \\
\hline\noalign{\smallskip}
HD\,45677     &  279 &       2.0  $\pm$ 0.9 & 0.3    $\pm$ 0.9  & 21.6 $\pm$ 2.0  & 1, 2 \\
HD\,259431    &  173 &    $-$2.4  $\pm$ 1.1 & $-$2.7 $\pm$ 0.9  & 19   $\pm$ 4.1  & 1, 3 \\
HD\,50138     &  392 &    $-$3.3  $\pm$ 0.6 & 4.1    $\pm$ 0.5  & 34.2 $\pm$ 2.0  & 1, 4 \\
CD$-$24\,5721 & 3500 &    $-$1.1  $\pm$ 2.7 & 2.0    $\pm$ 2.4  &                 & 5, 6 \\
CD$-$49\,3441 & 2000 &    $-$11.5 $\pm$ 1.7 & 18.0   $\pm$ 1.6  & 33 $\pm$ 2      & 6, 7 \\
HD\,85567     & 1500 &    $-$9.3  $\pm$ 0.6 & 6.2    $\pm$ 0.5  & 0 $\pm$2        & 1, 7 \\
HD\,98922     & 1190 &    $-$9.0  $\pm$ 0.4 & 0.8    $\pm$ 0.3  &                 & 1    \\
HD\,181615/6  &  546 &    $-$1.3  $\pm$ 0.2 & $-$6.3 $\pm$ 0.2  & 8.9 $\pm$ 0.9   & 1, 4 \\
MWC\,623      & 2000 &    $-$0.6  $\pm$ 2.1 & $-$3.9 $\pm$ 2.1  &                 & 5, 8 \\
\noalign{\smallskip}\hline
\hline
\end{tabular}
\tablebib{
$^{(1)}$ \citet{van07};
$^{(2)}$ \citet{eva67};
$^{(3)}$ \citet{gon06};
$^{(4)}$ \citet{wil53};
$^{(5)}$ \citet{mir03};
$^{(6)}$ \citet{hog00};
$^{(7)}$ \citet{mir01};
$^{(8)}$ \citet{zic01}
        }
\label{tab:pmrv}
\end{table*}


\section{Discussion}\label{sec:fscma}


Whether a young star can be in isolation from star-forming activity is an open 
issue.  \citet{gra93} suggested that the isolated B[e] stars with prominent infrared excess could be in the 
PMS phase.  However, most of the prototypical isolated Herbig stars are actually not far from SFRs.  
For example, UX\,Ori is only some 40~$\arcsec$ from the nearby 
luminous cloud NGC\,1788 and a group of T~Tauri stars.  Another well known example, HD\,163296 \citep{the85} 
is, when a large field was scrutinized, associated with dark clouds as a part of a prominent \mbox{H\,{\sc ii}} 
region and cloud complex \citep{lee11a}.   We note that HD\,163296 clearly shows elevated 
mid-infrared emission, signifying cold dust \cite[see Figure~4;][]{kra08a}.  
Even if isolated star formation could have been possible \citep{gri91,ale13}, 
a new-born star collapsing out of a molecular core should still keep a distributed dust, warm and cold, 
in the envelope; this is not the case for our targets. 
We therefore conclude that the eight stars reported here, namely 
HD\,50138, HD\,45677, CD$-$24\,5721, CD$-$49\,3441, MWC\,623, HD\,181615/6, HD\,85567, and HD\,98922, 
represent a well defined FS\,CMa sample.  
We note that the two newly identified FS\,CMa stars reported here, HD\,98922 and HD\,181615/6,
have IRAS colors $([25]-[12], [60]-[25]) = (-0.17,-0.64)$ and $ = (-0.49,-0.73)$, respectively, indeed 
consistent with the known ranges of this class \citep{mir07}.  Observationally the FS\,CMa stars 
show B[e] spectra and prominent excess emission in near-infrared, but only moderate at 
longer wavelengths.  They have evolved beyond the PMS phase, so the dust should not be surplus
from star-forming processes.



\citet{hil92} divided Herbig stars into three groups by their level of near- to mid-infrared excess relative to 
the stellar photospheric emission:~Group~I (moderate), Group~II (prominent), and Group~III (weak).  
Groups I and II show flat or rising SEDs toward mid-infrared, whereas Group~III sources have steep flux decrease 
toward longer wavelengths.  These authors classified HD\,259431, the only Herbig star in our sample, 
as a Group~I source.  In comparison, the eight FS~CMa stars we present here have comparable near-infrared excesses as Group~I Herbig stars, but with mid-infrared excess clearly even weaker than Group~III.    
In at least two stars in our sample, in HD\,45677 \citep{di08} and HD\,50138 (this work), the flux 
continues to decrease beyond far-infrared.   Millimeter measurements of HD\,45677, included in the SCUBA 
legacy catalogues \citep{di08}, indicated an emission size 14$\arcsec$, comparable to the SCUBA 
resolution FWHM$\sim 14\arcsec$ at 850~$\mu$m, that is, the emission was unsolved.  For our SMA observations 
of HD\,50138 at 1.3~mm, shown in Figure~\ref{sma}, with an angular resolution of $\sim 1\arcsec$, 
the emission was unresolved in either the image or the visibility domain.  The fall-off of the 
SEDs toward submillimeter wavelengths suggests a lack of cold grains, thus more compact dusty 
envelopes in size than those in Herbig stars.  Alternatively, photoevaporation would erode the outer 
parts of a circumstellar disk earlier than the inner part, thereby exacerbating the disk dispersal within 
timescales of a few Myr \citep{gor15}.  The classification scheme for Herbig stars based on infrared 
excess therefore is not directly applicable to the B[e] stars in our sample because of the different 
origins 
of circumstellar grains.  



Existence of thermonuclear products would provide a diagnostic of an evolved atmosphere.  
\citet{kra09} proposed that an evolved B[e] star should have a $^{13}$C-enriched equatorial environment 
because of the disk-forming winds in massive stars, locking the isotope in the form of molecules, 
for example, $^{13}$CO seen in high-dispersion K$-$band spectra.  \citet{oks13} and \citet{lie14} 
found $^{13}$CO emission in four FS~CMa stars, including MWC\,137, HD\,327083, GG\,Car, and 
Hen\,3$-$298, to justify the stellar seniority.  We found related literature only for 
two stars in our sample.
\citet{whe13} detected no $^{13}$CO emission in HD\,85567, but the spectral resolution of the data 
was not sufficient to dismiss the post-MS status.  In MWC\,623 $^{13}$CO is in absorption, which 
may be attributable to the cool supergiant companion \citep{lie14}.  In any case a sample of 
FS~CMa stars in star clusters would better firm up their evolutionary status.  
Recently, a few FS~CMa candidates were found as members of two young clusters, in 
Mercer~20 and Mercer~70 \citep{fue15}, though uncertainty in mid-infrared fluxes owing to cluster 
crowding requires further clarification.

Correlation between the level of infrared excess and intensity of Balmer emission lines has been known in Be 
stars \citep[e.g.,][]{fei81,net82,dac82,ash84,gor85,dac88,kas89,ker95,tou10}, 
in Herbig stars \citep{cor98,man06} and in T~Tauri stars \citep{cab90}.  For Be stars, those 
with H$\alpha$ and H$\beta$ both in emission exhibit large near-infrared excess \citep{lee11b}.   
All our targets belong to this group.  In comparison, those with H$\alpha$ in emission, 
but H$\beta$ and higher Balmer lines in absorption have moderate near-infrared excess, with $J-H$ 
and $H-K_{\rm s}$ up to 0.8~mag.  The Be stars with H$\alpha$ and H$\beta$ both in absorption, 
show little near-infrared excess, with the observed $J-H$ and $H-K_{\rm s}$ $\lesssim 0.2$.  

The correlation between the dust emission and emission lines supports the two-component model for  
FS\,CMa stars, one of an enhanced stellar mass loss, and the other of a dusty envelope \citep{zic03}.  
The copious hot plasma produces the prominent H$\alpha$ emission lines in the vicinity of the stars, 
and also the forbidden lines in the tenuous regions further away.  The expanding gas cools 
off and condenses to form dust grains, which reprocess starlight to give rise to the excessive 
infrared emission.  Because dust is formed out of stellar mass loss, proceeding inside out to the envelope, 
it is confined preferentially in the vicinity of the star.   
This is in contrast to a Herbig star which harbors existing large grains distributed 
on scales of the parental molecular core.

All isolated B[e] stars in our sample suffer comparatively small 
intrinsic 
extinction 
(Table~\ref{tab:targets}), despite the prominent near-infrared excess signifying existence of 
warm dust in quantity.  Possible explanations include an anisotropic dust distribution or 
ineffective dust extinction.
Dust formation in an asymmetric mass outflow, for example, as a consequence of colliding massive stellar winds, 
ejection in mass transfer in a binary system, or fast rotation to shed the atmosphere of a single star, 
results in a dusty region close in to the star and a relatively clear line of sight.  
The mass loss consequently gives rise to the P~Cygni profiles commonly seen in the B[e] spectra.

Alternatively, tiny grains are effective in thermal emission but otherwise not in attenuation 
of visible light.  Freshly condensed grains should be tiny in size.  Grain growth in an FS\,CMa envelope 
may proceed efficiently, but with ample supplies of tiny grains, such as nanosized particles, 
which are known to have anomalous specific heat capacity \citep{hul57,pur76}. 

It has been suggested that binarity may be responsible for the FS\,CMa phenomena \citep{pol06}, and 
dust is formed as a consequence of binary interaction.  Interestingly, among our targets, 
HD\,181615/16 and MWC\,623 have ascertained spectral types for both binary components, whereas 
HD\,45677, HD\,50138, HD\,85567, and HD\,98922 are found to be spectro-astrometric binaries, 
that is, the photocenter of a star is wavelength dependent \citep{bai06}.  
 
 
Dust formed out of stellar mass loss, therefore, has properties distinct from that in the 
interstellar medium or around a young star.  Caution must be taken, then, to derive dust contents 
in different environments based on 
interstellar 
grain properties.  The current paradigm is that 
most cosmic dust is produced in expanding cool atmospheres of 
post-main sequence stars.  Our work supports the notion that FS~CMa stars may serve as additional 
suppliers of dust in space \citep{mir07}.

\section{Conclusion}

We present a sample of eight B[e]-type stars that show prominent near-infrared excess, but steep flux drop 
toward mid-infrared and longer wavelengths, indicating the dusty envelopes being compact in size and lacking 
cold grains.  On the basis of their isolation from star-forming activity or young stellar populations, 
we argue that these stars have evolved beyond the PMS phase, so the grains must be
produced in situ, 
starting with very tiny sizes.  Anisotropic mass loss, or properties of tiny grains, lead to the phenomena 
observed in our targets.  The properties of dust, such as the size, composition, and spatial 
distribution in these isolated B[e] stars certainly deserve further investigation.  


\begin{acknowledgements}
Support for this work was provided by the the Ministry of Science and Technology of Taiwan through grant 103-
2112-M-008-024-MY3.
\end{acknowledgements}


\bibliographystyle{aa}
\bibliography{references}
\end{document}